\documentclass[twocolumn,aps,showpacs,prb,postscript,superscriptaddress]{revtex4-1}
\usepackage{amsmath,amssymb}

\usepackage{makecell}
\usepackage{subfig}
\usepackage{graphicx}
\usepackage{braket}
\usepackage{epstopdf}
\usepackage[usenames]{color}
\usepackage{hyperref}
\usepackage{float}
\usepackage{xcolor}
\usepackage{tikz}
\usepackage{comment}
\usepackage{setspace}
\usepackage{diagbox}
\usepackage[sort&compress]{natbib}
\usepackage{booktabs, multirow} 
\usepackage{soul}
\usepackage[normalem]{ulem}
\usepackage{changepage,threeparttable} 
\usetikzlibrary{positioning}

\begin{document}

\title{Dephasing and Decorrelation of Spins in a Disordered Environment}

\author{Shreyas Raman}
\affiliation{Department of Physics, Indian Institute of Science, Bangalore 560 012, India}
\author{Subroto Mukerjee}
\affiliation{Department of Physics, Indian Institute of Science, Bangalore 560 012, India}

\begin{abstract}
Dephasing of spins is a major roadblock to scaling up the size of quantum computing systems. We explore the possibility of utilizing highly disordered environments which are in the Many-Body Localized phase to arrest this dephasing. We embedded 2 `special' spins in such a highly disordered environment of Heisenberg spins to act as the target qubits and use the long-time value of the spin-spin correlator $\langle \vec{\sigma}_i \cdot \vec{\sigma}_j\rangle$ as an order parameter to quantify the transition between the thermal and MBL phases of this system. It is seen that the dephasing between spins, as encoded in this correlator, is impeded in a disordered environment when the system is fully localized. The order parameter yields a critical exponent, to characterize the transition between the thermal and MBL phases, that appears to be robust to changes in microscopic parameters of the system or the choice of pair of spins.
\end{abstract}

\maketitle

\section{Introduction}
Quantum computation, which promises to be faster than its classical analog in a range of applications\cite{Haystack} is based on the ability to store and process information encoded in a collection of qubits, which are in superpositions of states. An obstacle hindering the experimental implementation of a complex quantum computational algorithm is quantum decoherence, arising from the inevitable interaction between the system and its environment\cite{RePEc:nat:nature:v:404:y:2000:i:6775:d:10.1038_35005011, scully_zubairy_1997}. Further, entanglement of qubits via quantum gates\cite{Cirac, All-Optical}, which is an essential component of quantum computation, results in faster loss of computational fidelity\cite{Disentanglement}. Many quantum error-correction protocols require the decoherence to be below a certain threshold\cite{RevModPhys.75.715}. Thus, understanding and suppressing decoherence in qubits is of paramount importance to the field of quantum computation.

The problems of single-qubit decoherence and its dynamical control have been investigated thoroughly\cite{8(2), 9(2)} and extended to multipartite decoherence control\cite{10(2)}. The specific strategy to protect quantum information from decoherence depends on the properties of the noise. When the coupling of different qubits is uncorrelated, such as in the case of superconducting circuits, an effective strategy is dynamical decoupling (DD), achieved by periodically applying a series of pulses to the system to refocus the system-environment evolution\cite{4(5), 5(5), 6(5)}. Recent experiments have demonstrated significant improvement of coherence times resulting from pulsed DD\cite{9(5), 10(5), 11(5), 12(5), 13(5)}. However, implementation of this method is challenging for systems with fluctuating noise. Further, combining the decoupling scheme with gate operations so that decoherence is suppressed is challenging. 

The above mentioned methods have primarily focused on tuning the system under consideration in an attempt to suppress decoherence. On the other hand, there exists the possibility of modifying the environment parameters to do the same. In this work, we propose tuning the environment to suppress decoherence by introducing disorder into the environment. The motivation for this strategy comes from noting that quantum systems reach thermal equilibrium starting from an initial state through decoherence\cite{20(3)}. In such systems, we typically find that the microscopic information associated with any initial state is lost as the system evolves towards equilibrium. Information about any initial state spreads over the entire system, rendering local measurements incapable of any useful interpretation\cite{1(3), 2(3), 3(3)}. Disorder-induced localization is a phenomenon that was first shown by Anderson in 1958 in non-interacting systems\cite{1(4)}. This localized phase has been observed for systems of non-interacting phonons, photons and matter waves\cite{4(4), 6(4)}. Recent work further shows that localization persists even in the presence of interactions between particles, establishing the Many-Body-Localized (MBL) phase as a robust, non-ergodic phase of quantum matter at finite temperatures\cite{7(4), 8(4), 9(4), 10(4), 11(4), 5(3), 7(3), 8(3), 9(3), 10(3),11(3), 12(3), 13(3), 14(3), 15(3), 16(3), 17(3), 18(3), 19(3), 20(3), 21(3), 22(3), 24(3), 25(3), 26(3), 27(3), 28(3), 30(3), 31(3)}. Strong disorder, giving rise to localization can prevent thermalization, leading to ``memory'' of initial states even at late times. From an information theoretical perspective, the ability of localized systems to avoid thermalization suggests their potential use as memory resources. The presence of strong interactions leads to a logarithmic growth of entanglement entropy, manifesting as an intrinsic, slow decoherence mechanism\cite{10(3), 11(3), 12(3), 13(3), 14(3), 20(3)}, which has been experimentally observed\cite{30(4), 31(4)}. A recent numerical study of the dynamics on the ergodic side of the transition in a Floquet model of many-body localization in large systems sizes has revealed clear slow dynamics for a given system size. This is reflected in a stretched exponential decay of the autocorrelation function(equal position, unequal time correlator) and subballistic spreading of the entanglement entropy; with the dynamics consistently speeding up with increasing system size and decreasing disorder strength\cite{Lezama_2019}.

The ergodic to MBL transition has been a topic of intense study for many years. However, the exact nature of the transition is still not well understood\cite{Alet_2018, Khemani, con10, con9}. A part of the problem lies in the fact that it appears not to be describable in terms of symmetry breaking with an appropriate order parameter like a conventional continuous transition. Thus, it is not obvious whether the transition can be characterized by critical exponents (and if so, by how many) that are robust to microscopic details. In fact, the ergodic to MBL transition is ``dynamic'' in the sense that it separates a phase (the ergodic phase) which dynamically equilibrates starting from an arbitrary initial condition from one which does not (the MBL phase). Since the dynamics is completely determined by the spectrum of energy eigenvalues and eigenstates, the ergodic to MBL transition is captured in the change of properties of the eigenstates and the distribution of eigenvalues across the transition. Even in the absence of any obvious symmetry breaking, attempts have been made to characterize the transition in terms of different order parameters. Examples of such order parameters are the scaling of the entanglement entropy with system size\cite{Alet_2018}, the eigenstate to eigenstate fluctuation of physical observables\cite{Deng, Xiaopeng}, the density imbalance at long times starting from an initial non-zero value\cite{31(3)}, the fluctuation of entanglement entropy from eigenstate to eigenstate, one disorder realization to another and from cut to cut to within a single system\cite{Khemani}. However, the nature of the transition and even whether it is continuous or not remains inconclusive from these studies.

Several studies have also focused on obtaining a diverging correlation length to characterize the transition. Studies on finite-sized systems\cite{18(3), 21(3), Khemani} of microscopic models with quenched randomness have produced correlation length exponents $\nu$ that violate the Chayes-Chayes-Fisher-Spencer (CCFS) bound $\nu > 2$\cite{Xiaopeng} suggesting that they are not in the proper scaling regime to obtain an accurate value of $\nu$. A complementary approach has utilized renormalization group methods for phenomenological models with coarse-grained interacting thermal and insulating regions \cite{24(3), Goremykina, Thiery}. The most fruitful of these has been based on the avalanche mechanism \cite{Thiery} have produced exponents consistent with the CCFS bound and in fact, even $\nu=\infty$ suggesting that the transition is Kosterlitz-Thouless like\cite{Kosterlitz}.

Numerics on finite-sized systems have so far not found rare thermal regions required to seed avalanches. This has been attributed to the insufficient sizes of these systems. It has been argued that the ``true'' MBL to thermal transition in the thermodynamic limit which occurs via avalanches is at a disorder strength significantly larger than obtained from finite-size numerics\cite{con10}. Additionally, the ``transition'' observed in numerical studies has been claimed to be a crossover (albeit a fairly sharp one) from a finite-size MBL phase destabilized by system-wide many-body resonances to a thermal phase\cite{con10, con9}. It should be emphasized that the true MBL transition remains out of reach numerically and also experimentally owing to limitations of system size and time scales. Thus, it is only the finite-size crossover, which can be studied in microscopic models without an appeal to phenomenology and it is still not well understood. The existence of the MBL phase in the thermodynamic limit has also been questioned with no proper evidence currently available to resolve the issue \cite{Jan, con2, con4}.

In this paper, we focus on the sharp MBL to thermal crossover accessible to numerics on finite-sized systems. To be consistent with the prevailing terminology, we refer to this crossover as a ``transition'' with the understanding that it is not expected to be one in the thermodynamic limit. The particular quantity we study is the equal-time correlator of a pair of ``special'' spins embedded in the rest of of the system, which can be in the thermal or the MBL phase depending on the strength of the disorder. These special spins, which are not subjected to disorder themselves, interact with the rest of the system. Our system is thus a very simple model of two qubits in an environment whose properties can be tuned with changing disorder strength. We show that the asymptotic value of the above-mentioned correlator is suppressed with increasing disorder strength. Further, we argue that this quantity can act as an ``order parameter'' for the transition and numerically extract a critical exponent that appears to be fairly robust to changes in the microscopic parameters.

This paper is organized as follows : In section II, we introduce the system under consideration along with its ratio of successive level spacings\cite{Alet_2018, 7(3)}, to identify the transition between the thermal and MBL phase. Further, we define the spin-spin correlator, which we employ as a diagnostic to quantify decoherence. In section III, we demonstrate that the correlator serves as an order parameter for the transition and obtain a critical exponent from the same for the transition for different system parameters. The data obtained has been tabulated in Section IV.

\section{Spin-Spin Correlator in a Disordered Environment}
Consider the following one-dimensional lattice model, the spin-$\frac{1}{2}$ Heisenberg chain in a random magnetic field 
\begin{equation}
    \hat{\mathcal H} = \sum_i \left(\hat \sigma_i^x \hat \sigma_{i+1}^x + \hat \sigma_i^y \hat \sigma_{i+1}^y + \hat \sigma_i^z \hat \sigma_{i+1}^z\right) +\sum_i h_i \hat \sigma_i^z
\end{equation}
where the system has periodic boundary conditions, $\sigma^{x, y, z}$ are the Pauli spin matrices and the $h_i$ are drawn from a uniform distribution  $\left[-h, h\right]$. We first use the Jordan-Wigner transformation to convert this to a model of hard-core bosons for computational fidelity. With some re-scaling, this can be written as a simple tight-binding system of hard-core bosons with nearest neighbour hopping and interaction with random on-site potential, i.e 
\begin{equation}
    \hat{\mathcal H} = \sum_i \left(\hat b_i^\dagger \hat b_{i+1} + \hat b_{i+1}^\dagger \hat b_i + 2\hat n_i \hat n_{i+1}\right) + \sum_i h_i \hat n_i 
\end{equation}
where $b$ and $b^\dagger$ are the hard-core boson annihilation and creation operators respectively and $n_i$ is the occupation of site $i$. Note that this Hamiltonian preserves boson number, just like the original Hamiltonian preserves total spin. The numerical analysis of this paper will be done on a system with equal number of up and down spins, or in terms of bosons, a system with half-filling.\\\\ \textit{Eigenvalue Statistics :} In the thermal phase, the eigenvalues follow the statistics of Random Matrix Theory. As the Hamiltonian has only real entries, this will be the Gaussian Orthogonal Ensemble (GOE). A useful computational quantifier to verify this is the ratio of consecutive level spacings \cite{Alet_2018, 7(3)}
\begin{equation}
    r_n = \frac{\min\left(E_n - E_{n-1}, E_{n-1} - E_{n-2}\right)}{\max \left(E_n - E_{n-1}, E_{n-1} - E_{n-2}\right)}
\end{equation}
Its average value in the GOE can be computed as $\langle r \rangle_\text{GOE} \approx 0.5307$. On the other hand, in the MBL phase, the eigenvalues are non-correlated and hence they follow Poisson statistics. The average value in the Poisson distribution can be computed as $\langle r \rangle_\text{Poisson} \approx 0.386$. For this system, we compute the average ratio by averaging over disorder realizations.

\subsection{Modified System with special spins}
We explore the viability of the spin-spin correlator between 2 `special' spins as an order parameter for the phase transition between the thermal and MBL phases. The 2 special spins are set such that they do not have a random on-site potential, while every other spin does. In terms of bosons, this is equivalent to picking 2 special sites($i$ and $j$), which do not have any random on-site potential. The Hamiltonian is : 
\begin{align}
    \hat{\mathcal H} &= \sum_k \left(\hat b_k^\dagger \hat b_{k+1} + \hat b_{k+1}^\dagger \hat b_k + 2\hat n_k \hat n_{k+1}\right) + \sum_{k; k \neq i, j} h_k \hat n_k \\ \hat{\mathcal H}(V) &= \sum_k \left(\hat b_k^\dagger \hat b_{k+1} + \hat b_{k+1}^\dagger \hat b_k + V\hat n_k \hat n_{k+1}\right) + \sum_{k; k \neq i, j} h_k \hat n_k 
\end{align} 
where $V$ is a parameter introduced as a microscopic system parameter - the Interaction Potential Strength - that can be tuned (explained later). When not mentioned, $V = 2$.

\begin{figure}[H]
    \centering
    \begin{tikzpicture}
    \node (img1){\includegraphics[width = 0.45\textwidth]{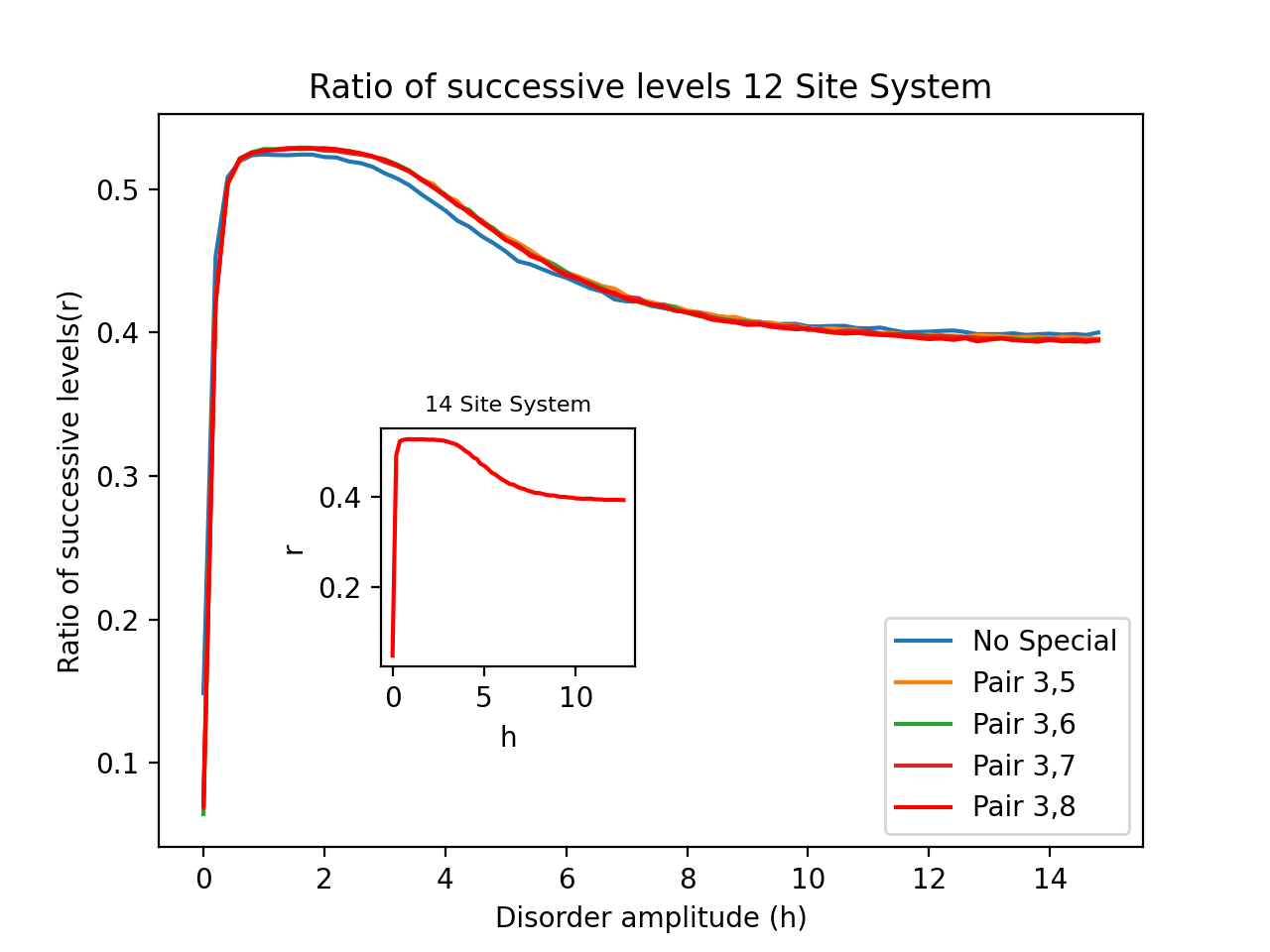}};
    \end{tikzpicture}
    \caption{Ratio of successive levels averaged over 2000 disorder realizations. Ratio plotted for different choices of special spins (including no special pair). Inset : Ratio of successive levels for a 14 site system for the spin pair 3, 8.}
    \label{fig:1}
\end{figure}

Comparing the ratio of successive levels of the original system with the modified system (FIG. \ref{fig:1}) confirms that in terms of level-spacing statistics, there is no major change in the system. The spin-spin correlator we want to evaluate is $\langle \vec{\sigma_{i}} \cdot \vec{\sigma}_{j}\rangle$. In terms of boson operators, this is : \begin{equation}
    \langle \vec{\sigma_i} \cdot \vec{\sigma_j} \rangle \equiv 4n_{i} n_{j} - 2n_i - 2n_j + 1 + 2\langle b_i^\dagger b_j \rangle + 2\langle b_j^\dagger b_i \rangle
\end{equation}

\begin{figure}[H]
    \centering
    \begin{tikzpicture}
    \node (img1){\includegraphics[width = 0.45\textwidth]{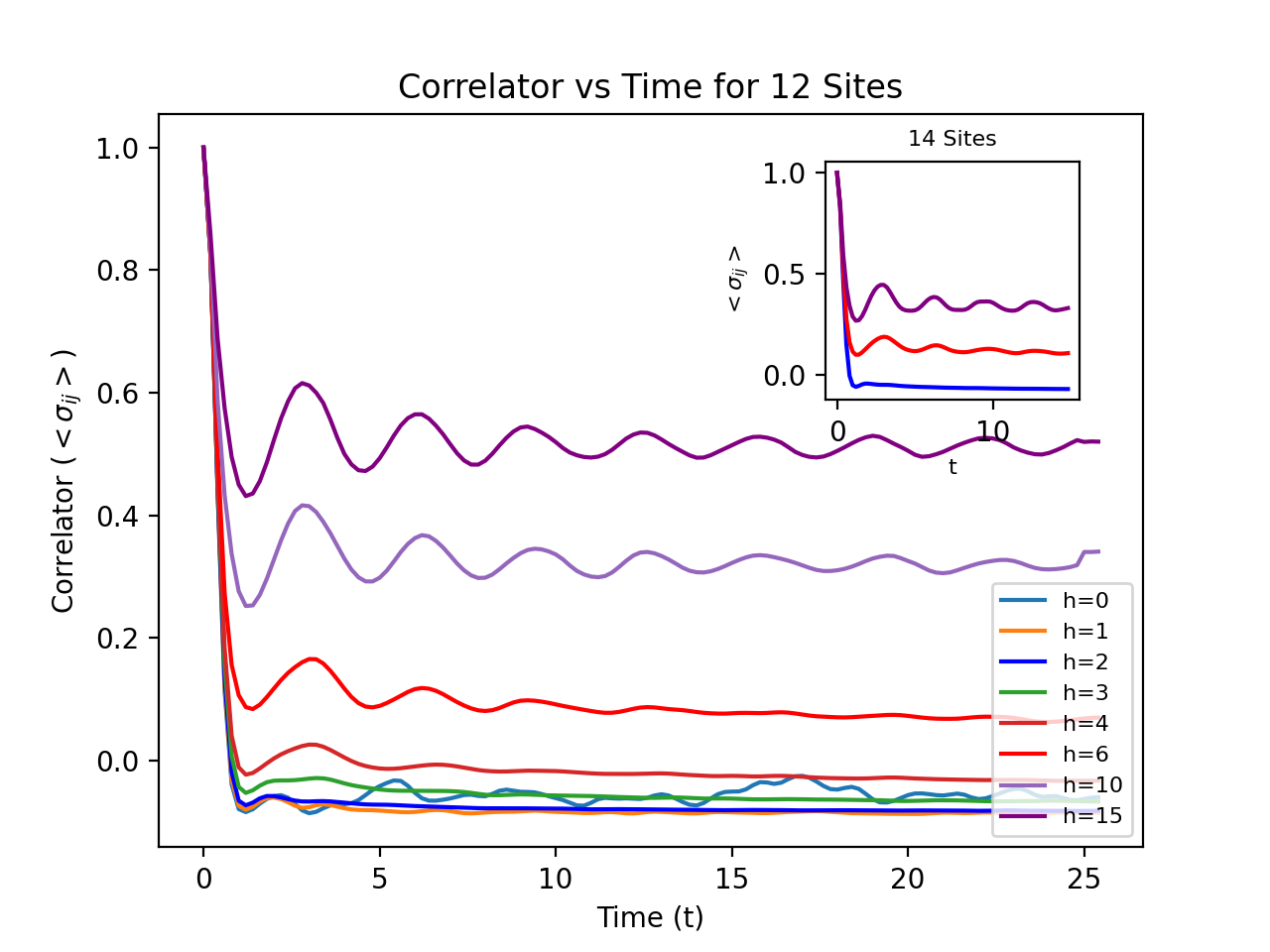}};
    \end{tikzpicture}
    \caption{Correlator vs. Time for $d = 5$ and for different values of disorder amplitude; averaged over 2000 disorder realizations and 210 initial configurations. Inset : Correlator vs Time for 14 Site system for $h = 2,6,10$, averaged over 1500 disorder realizations and 792 initial configurations}
    \label{fig:2}
\end{figure}

The system is initialized such that at time $t = 0$, it is in a product state where both the special spins are initialized to `up'. We then average over all possible initial product states that have this property. This ensures the initial correlator is exactly 1. The correlator as a function of time was plotted for the system, averaged over disorder realizations and initial configurations(FIG. \ref{fig:2}). A fixed distance between spin pairs was chosen, defined as $d \equiv |j - i|$.

\begin{figure}[htbp]
    \centering
    \begin{tikzpicture}
    \node (img1){\includegraphics[width = 0.45\textwidth]{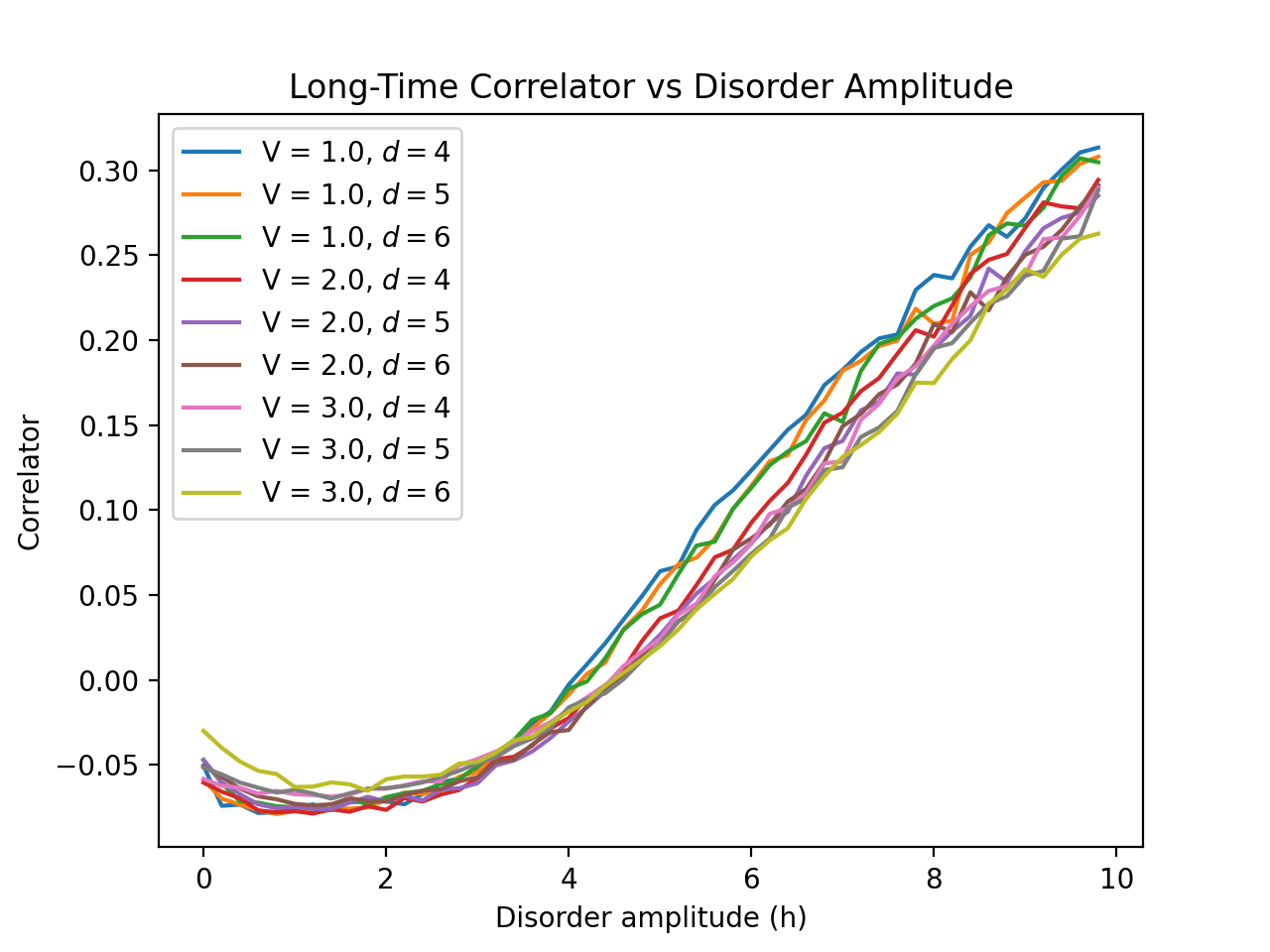}};
    \end{tikzpicture}
    \caption{Long-Time value of the Correlator vs. Disorder Amplitude for a 12 site system; for different values of interaction potential and choice of spin pair. The trend shown by the long-time correlator motivates its use as an order parameter to quantify this transition and subsequent extraction of a power law close to the transition point.}
    \label{fig:3}
\end{figure}

The trend shown by the long-time value of the correlator indicates that it could be a good order parameter for characterizing the phase transition between the thermal and MBL phases. For small disorder strength $h$, the long-time correlator tends to the same value, characteristic of the system thermalizing. In the infinite system size limit, this value tends to 0. From here, as $h$ increases, the long-time correlator also increases. The trend for the long-time correlator as a function of disorder strength is shown in FIG. \ref{fig:3}.

\section{Transition Points and Critical Exponents}

To find the transition point of an $N$ site system,
we found the intersection point of the ratio of successive
levels between an $N$ site system and an $N-2$ site system(FIG. \ref{fig:4}). All the other parameters of the system
were kept the same. 

A power law was extracted from the trend of long-time correlator vs. disorder strength close to the transition point. To do this, we assume the relation is of the form (to the right of the transition point) : 
\begin{equation}
    \langle \vec{\sigma}_i - \vec{\sigma}_j \rangle - \langle \vec{\sigma}_i - \vec{\sigma}_j\rangle_c = A(h - h_c)^\beta
\end{equation}
where the subscript $c$ stands for the value at the critical (transition) point.

The robustness of this power-law exponent can be checked against two major parameters : 
\begin{enumerate}
    \item Choice of spin pair
    \item Interaction Potential
\end{enumerate}

\subsection{Choice of Spin Pair}
As shown by FIG \ref{fig:1}, the physics of the system does not depend on the choice of spin pair and transition point of the system is independent of choice of spin pair. 

\subsection{Interaction Potential}
\begin{figure}[H]
    \centering
    \begin{tikzpicture}
    \node (img1){\includegraphics[width = 0.45\textwidth]{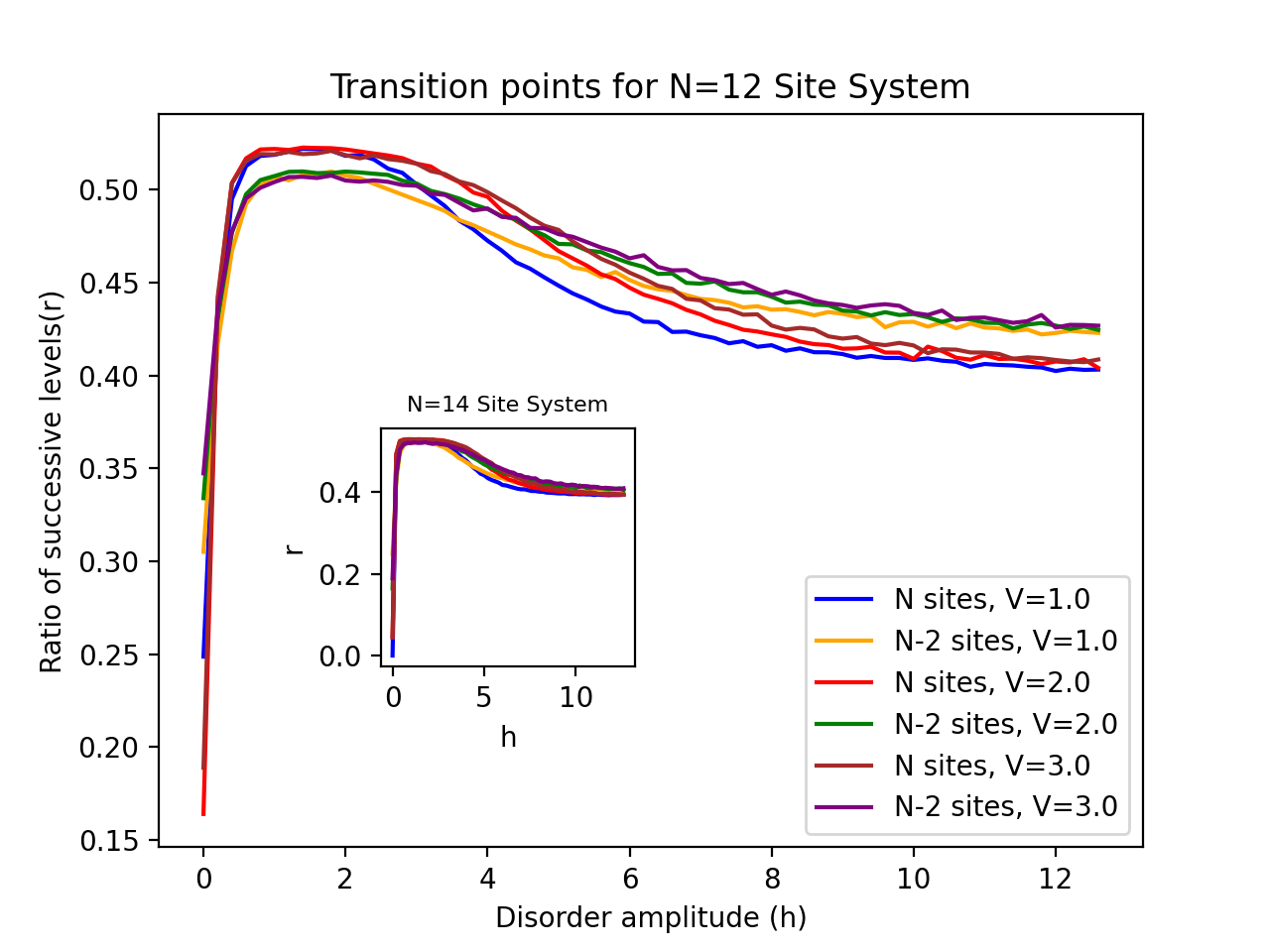}};
    \end{tikzpicture}
    \caption{Ratio of successive levels showing the transition points for different values of interaction potential. Inset : Ratio of successive levels showing the transition points for a 14 site system.}
    \label{fig:4}
\end{figure}

\begin{figure*}[t]
\centering
\begin{tikzpicture}
\node (img1){\includegraphics[width=0.5\textwidth]{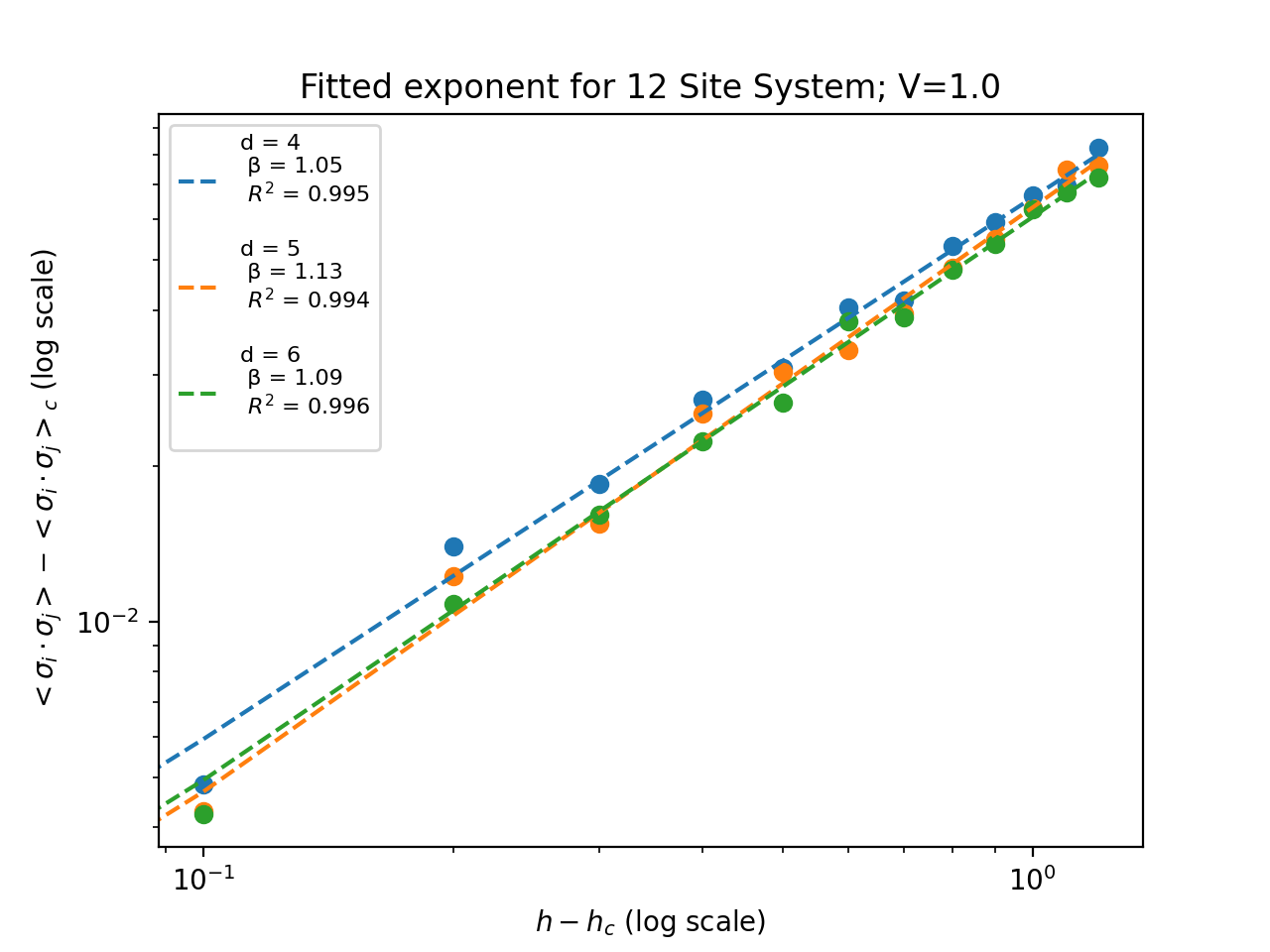}};
\node (img2)[right= of img1, node distance=0cm, xshift=-1.7cm]{\includegraphics[width=0.5\textwidth]{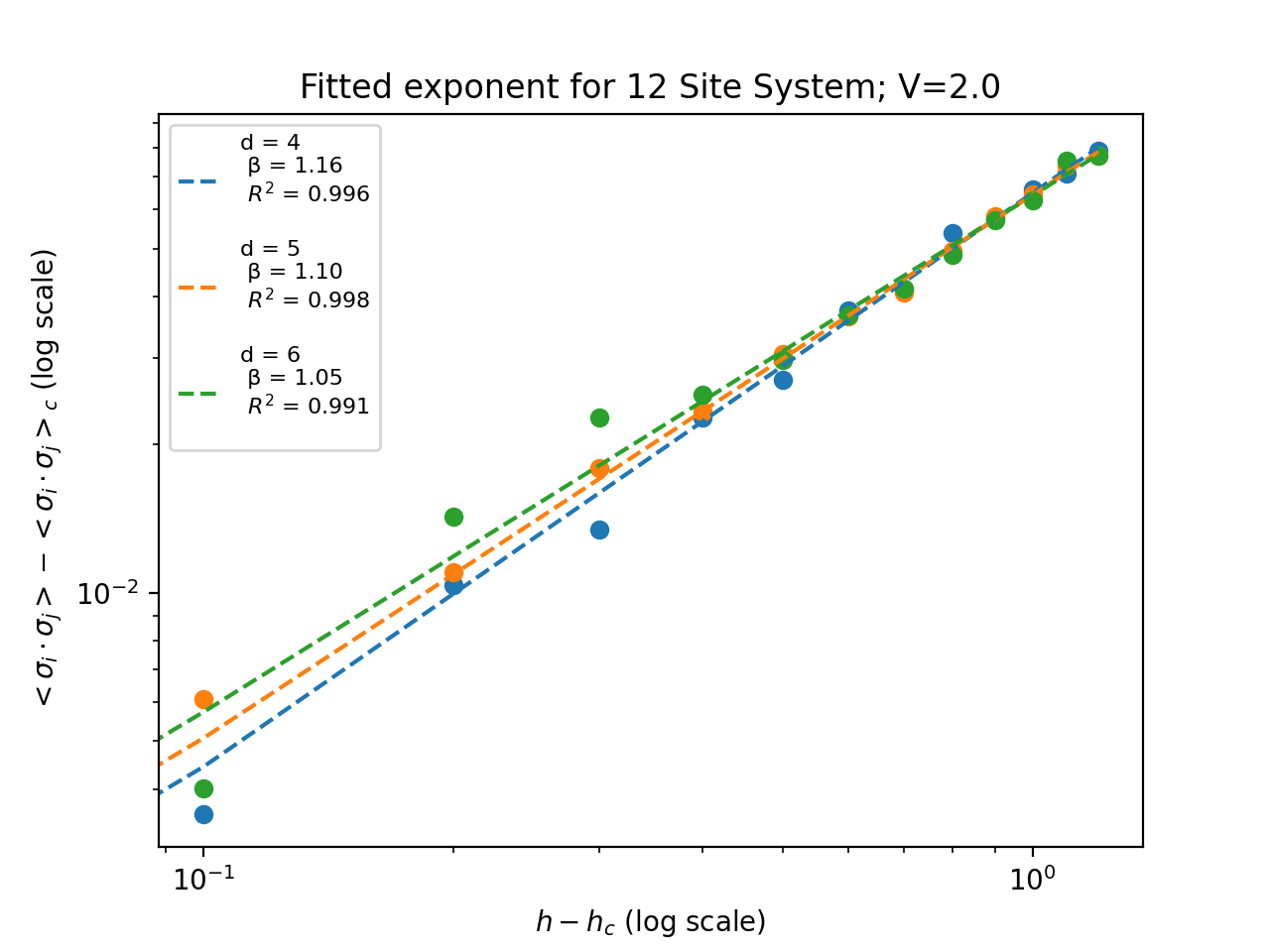}};
\node (img3)[below= of img1, node distance=0cm, yshift=1.0cm,xshift=0cm]{\includegraphics[width=0.5\textwidth]{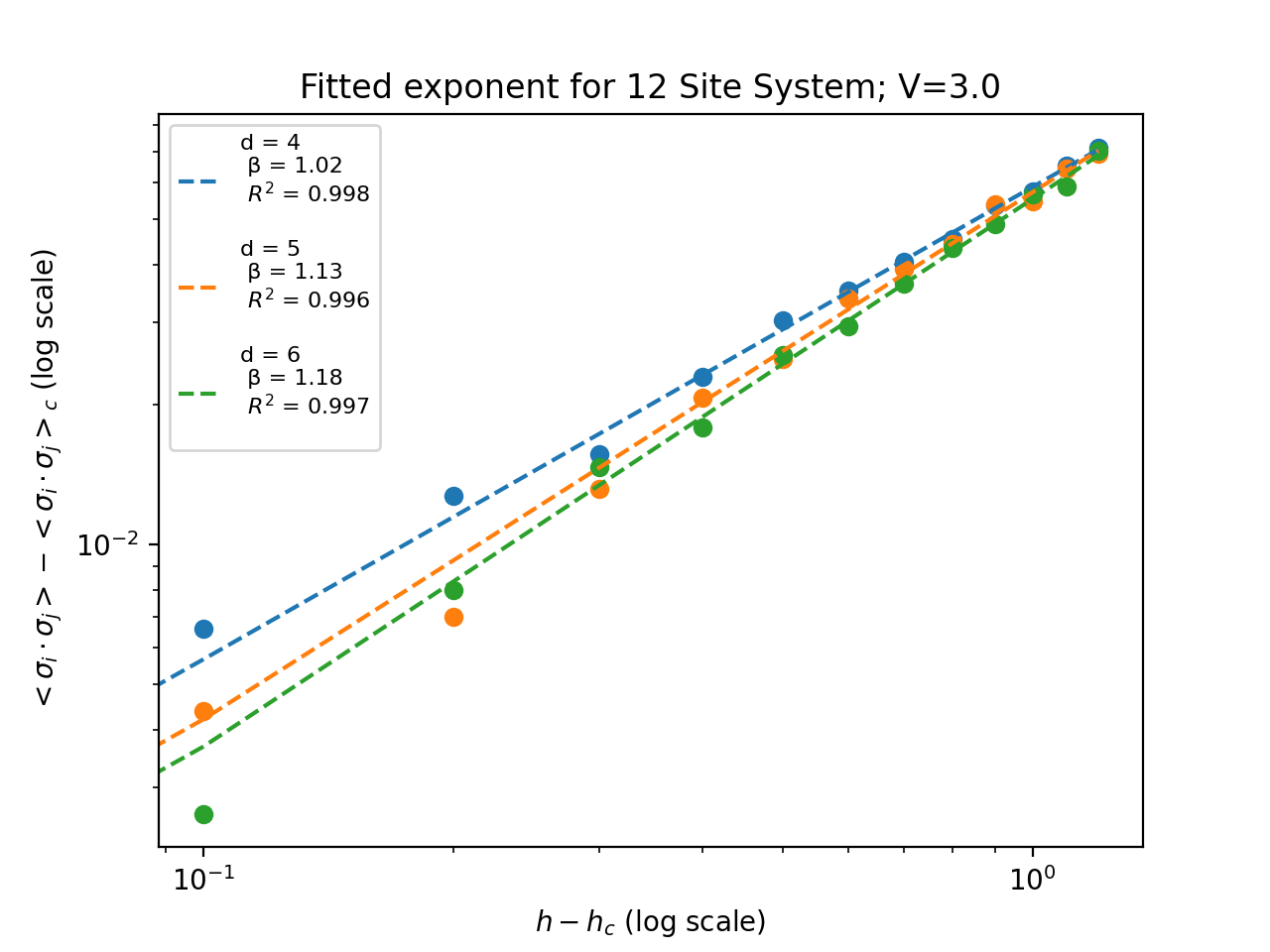}};
\node (img4)[below= of img2, node distance = 0cm, yshift=1.0cm,xshift=0cm]{\includegraphics[width=0.5\textwidth]{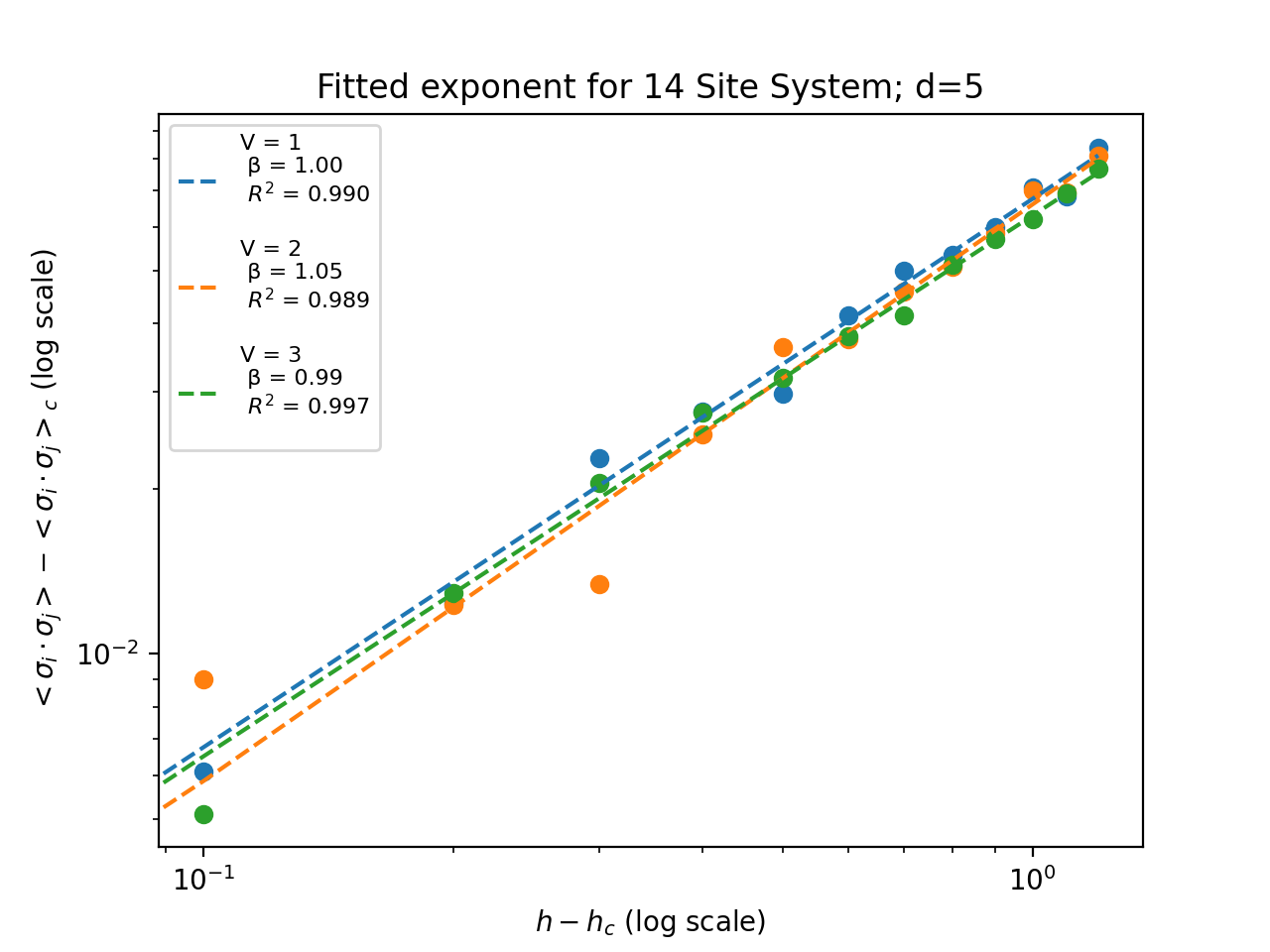}};
\end{tikzpicture}
\caption{Close to the transition point, it was assumed that the long-time correlator follows a power law as a function of disorder strength of the form $\langle \vec{\sigma}_i \cdot \vec{\sigma}_j\rangle - \langle \vec{\sigma_i}\cdot \vec{\sigma}_j\rangle_c \sim (h - h_c)^\beta$. The plot shows the fitted Critical Exponent close to the transition point for different values of interaction potential and choice of spin pair. The exponent obtained has an average value that seems to tend towards $1$ with increasing system size and is robust to system parameters (choice of spin pair and interacting potential).}
\label{fig:5}
\end{figure*}

This far in the paper, we have mainly discussed $V = 2$. As we explore the robustness of the long-time value of the correlator as an order parameter, we turn to changing $V$ as a microscopic parameter that changes the physics of the system. As the form of the Hamiltonian changes on changing the interaction potential $V$, the transition point of the system also changes. In order to obtain the new transition point, we plot the ratio of successive levels for 2 system sizes with the changed potential and locate the new intersection point (FIG. \ref{fig:4}).

A power law is fitted to the long-time value of the correlator to the right of the transition points as obtained from the intersection in FIG. \ref{fig:4}. The exponent values for some values of $V$ are obtained in FIG. \ref{fig:5} and the data has been tabulated in Table \ref{tab:1} and Table \ref{tab:2}.
\section{Data}

\begin{table}[H]\centering
\caption{Exponent $\beta$ cross-referenced by Potential and distance between spin pairs for 12 Site system}\label{tab:1}
\scriptsize
\begin{tabular}{|c|c|c|c|c|}\toprule\hline
\backslashbox{$V$}{$d$} &\textbf{4} &\textbf{5} &\textbf{6} &\textbf{Average}\\\hline
\textbf{0.5} &1.1 &1.15 &1.15 &1.13 $\pm$0.03 \\
\textbf{1.0} &1.09 &1.08 &1.09 &1.09 $\pm$0.01 \\
\textbf{1.5} &1.07 &1.12 &1.04 &1.08 $\pm$0.04 \\
\textbf{2.0} &1.11 &1.12 &1.05 &1.09 $\pm$0.04 \\
\textbf{2.5} &1.1 &1.07 &1.09 &1.09 $\pm$0.01 \\
\textbf{3.0} &1.02 &1.09 &1.18 &1.1 $\pm$0.08 \\\hline
\textbf{Average} &1.08 $\pm$ 0.03&1.11 $\pm$ 0.03 &1.10 $\pm$0.05 &1.10 $\pm$ 0.04 \\\hline
\bottomrule
\end{tabular}
\end{table}
\begin{table}[H]\centering
\caption{Exponent $\beta$ for different values of interacting potential for 14 site system with $d = 5$}\label{tab:2}
\begin{tabular}{|c|c|}\toprule \hline
     \textbf{Potential}($V$)& \textbf{Exponent}($\beta$) \\\hline
    0.5 & 1.03\\ 1.0 & 1.00\\ 1.5 & 1.05\\ 2.0 & 1.05\\ 2.5 & 1.00\\ 3.0 &0.99 \\\hline \textbf{Average} &  1.02 $\pm$ 0.03\\\hline
    \bottomrule
\end{tabular}
\end{table}

\section{Conclusion and Discussion}
From the calculations performed on a modified Heisenberg spin chain with a pair of special spins, it appears that the equal time spin-spin correlator of these spins can serve as an order-parameter for the transition of the rest of the system, from the thermal to MBL phase. The asymptotic value of this quantity is non-zero in the MBL phase of the environment and equal to zero (modulo finite-size effects) in the MBL phase. This clearly demonstrates that the decoherence of these spins is impeded in the presence of a disordered environment, which can find use in the design of quantum memories and quantum computing platforms. Interestingly, we find that the correlator yields a critical exponent that appears to have a value close to unity irrespective of the choice of parameters of the Hamiltonian or the distance between the special spins. This appears to justify the use of the correlator as an order parameter and hints at underlying universal features of the transition. The existence of such a universality would be interesting since, as mentioned in the introduction, it has been recently argued that the transition being studied here is, in reality only a sharp crossover. Whether the apparent robustness of the exponent obtained survives to higher system sizes is currently a matter of conjecture because of numerical limitations. Nevertheless, our observation is still significant given that all existing numerical studies on microscopic systems and indeed experimentally accessible systems appear to only be able to probe the finite-sized MBL transition and not the one that has been claimed to occur in the thermodynamic limit.

\section{Acknowledgments}
SM thanks the Department of Science and Technology, Govt. of India for funding for this work through the Quantum Information Science and Technology (QuST) scheme.

\end{document}